# Can secondary nucleation exist in ice banding of freezing colloidal suspensions?

Jiaxue You[1], Jincheng Wang[1], Lilin Wang[2], Zhijun Wang[1*], Junjie Li[1] and Xin Lin[1*]

1-State Key Laboratory of Solidification Processing, Northwestern Polytechnical University, Xi'an 710072, P. R. China

2-School of Materials Science and Engineering, Xi'an University of Technology, Xi'an 710048, P. R. China

**Abstract:** The formation mechanism of ice banding in the system of freezing colloidal suspensions, which is of significance in frost heaving, ice-templating porous materials and biological materials, still remains a mystery. Recently, the theory of secondary nucleation and growth of ice has been proposed to explain the emergence of a new ice lens. However, this theory has not been quantitatively examined. Here, we quantitatively measured the initial interfacial undercooling of a new ice lens and the nucleation undercoolings of suspensions. We found that the interfacial undercooling can not satisfy the nucleation undercooling of ice and hence disprove the secondary nucleation mechanism for ice banding.

Ice banding (or ice lens) is a familiar pattern of ice segregation in concentrated colloidal suspensions systems, that features alternating macroscopic layers of ice and colloids transverse to the temperature gradient [1-3]. These transverse segregated ice lenses (shown in Figure 1) are of significant importance, because they are closely related to the formation of frost heaving [4-6], mechanical properties of ice-templating porous materials [7, 8] and tissues of biological materials [9].

The microscopic patterns of ice banding have been extensively investigated since Taber's experiments on frozen soils in 1929 [10, 11]. The most widely accepted theory for ice lens formation is the "rigid ice" model, which treated the suspensions as a rigid or elastic porous matrix [12-14]. This model suggested that ice lensing in rigid-ice formulations requires the existence of a "frozen fringe", a region of partially

---

frozen suspensions extending beyond the warmest ice lens wherein the succeeding ice lens is initiated [2, 15]. However, using Raman spectroscopy analysis, Watanabe et al. could not detect a pore-ice-bearing fringe in front of ice lenses and thus suspected the "rigid ice" model [15]. Afterwards, a disequilibrium mechanism based on particle trapping was presented to explain the ice banding in rapid solidification of colloidal suspensions [16] (i.e. >100μm/s), drawing on an analogy with solute banding in rapid alloy solidification [1]. However, this model ignores effects of the concentrated particle layer ahead of the interface though it is rigorous in mathematics.

Recently, the theory of secondary nucleation of ice has been proposed to explain the emergence of a new ice lens, corresponding to the engulfment of a layer of particles. A new ice nucleus will appear, with the help of a nucleator, in front of the undercooled freezing interface. This theory is based on the particulate constitutional supercooling (PCS) caused by concentrated particles in front of the advancing freezing interface and without the requirement of a frozen fringe [3, 6, 17]. This theory seems to be an correct way to clarify the formation of ice banding in the system of freezing concentrated colloidal suspensions. Nevertheless, we currently found that the PCS is usually too small and has no effect on pattern formation if the effect of solutes is dominant [18]. On the other hand, nucleation requires a relative large undercooling because of the fierce competitions between Gibbs free energy of bulk and interfacial free energy at the initial stage of phase transformation [19-22]. Moreover, the temperature at the growth surface of the ice lens was estimated to be -0.06 $^0$C, which may be insufficient to provide enough nucleation undercooling [15]. These imply that the theoretical model of secondary nucleation may be not correct.

Currently, the mechanisms of frozen fringe and secondary nucleation are two popular viewpoints for how the new ice lens forms. However, both of the two have imperfections. The frozen fringe, just a hypothesis, has never been observed in the precise experiments [15, 23]. The secondary nucleation, even the help of nucleator, is difficult. A "geometrical supercooling" (i.e., PCS) was proposed to supply the undercooling of secondary heterogeneous nucleation [6]. However, the magnitude of geometrical supercooling is extremely small [18], which implies the secondary heterogeneous nucleation may not exist. Here we focus on the latter and a quantitative examination for the secondary nucleation theory is needed, which will present deep insights of ice lensing.

In this paper, we carried out a careful examination on the secondary nucleation theory via directly measuring the interfacial undercooling of a new ice lens and the nucleation undercooling of freezing colloidal suspensions. By comparing the nucleation undercooling with the interfacial undercooling, the reasonability of secondary nucleation theory in freezing colloidal suspensions is revealed.

In the experiments, the α−alumina powder with a mean diameter d=50nm and a density of 3.97 g cm$^{-3}$ was utilized (Wanjing New Material, Hangzhou, China, ⩾ 99.95% purity, monodispersity). The alumina suspensions were prepared by using HCl (hydrogen chloride) and deionized water as the solvent following Ref.[2]. Also the stable dispersity of alumina suspensions has been confirmed in Ref.[2]. The initial volume fraction of particles is $\phi_0$=9.74% (30wt%). The Bridgman freezing setup and experimental procedure have been described in Ref. [24]. During directional freezing, the temperature gradient keeps as G=7.23K/cm and the pulling speed is V=16μm/s. Figure 1 shows the ice banding formed as the particles were engulfed periodically by the advancing freezing interface. The dynamic formation process of this periodical ice banding is also presented in the Movie S1 (Supplementary Information).

As to the formation mechanism of ice banding, the emergence temperature of a new ice lens is the essential feature. So far, the interface undercooling that a new ice lens initially emerges has been conjectured theoretically and without unanimous conclusion [12, 14]. Although some experimental facilities have been applied to investigate the formation of ice banding and ice lens [2], the interface temperature of a new ice lens can hardly be captured in previous experimental investigation of ice lens, due to the big size of the Hele-Shaw cell (380 × 100 × 3 mm) and the large gap between the heating and cooling zones (60mm) [2, 25]. The experimental apparatus used here is exquisite and can be used to in situ observe the formation of ice banding and accurately determine the interfacial temperature that a new ice lens initially emerges [24].

Figure 2 shows the measured temperature of a new ice lens through the interface position difference between the supernatant (left cell of Fig. 2) and the suspensions (right cell of Fig. 2) within a microscopy. A linear thermal gradient is built across the upper and the bottom ends of the cell, which are the heating zone and the cooling zone respectively (indicated in Fig.1). Accordingly, the temperature measurement is converted into distance measurement in the thermal gradient platform. The position of solid/liquid interface in the supernatant cell (red dot line in Fig. 2) is slightly higher than that of emergence of a new ice lens (blue dot line in Fig. 2). The discrepancy of the solid/liquid interface positions between the supernatant and the new ice lens is 13.83μm, corresponding to a undercooling of 0.01 $^0$C under G=7.23K/cm. This indicates that the interfacial temperature of the supernatant is only slightly higher than that of the new formed ice lens. The interfacial temperature of the supernatant was calibrated as -0.03 $^0$C, by comparing with freezing deionized water (0 $^0$C), under G=7.23K/cm and V=16μm/s. Based on the interfacial temperature of the supernatant, the measured temperature of a new ice lens is around -0.04 $^0$C. This measurement is consistent with the theoretical predictions (around 0 $^0$C) in Refs.[26, 27] and experimental data (-0.06 $^0$C) in Ref.[15]. In addition, Movie S1 also clearly shows that the interface position that a new ice lens initially emerges is very near the interface position of the supernatant (-0.03$^0$C).

Since the undercooling (-0.01 $^0$C) that a new ice lens emerges is very small, it might not afford the nucleation undercooling in the secondary nucleation theory. The nucleation undercooling for the alumina suspensions is an important factor determining the success or failure of the mechanism of secondary nucleation. In the literatures reviewed, the nucleation undercooling of suspensions has rarely been gauged in the fields of freezing colloidal suspensions. Nevertheless, there are many experiences about the measurements of nucleation undercooling in alloy solidification [19, 20]. In the refrigeration of the liquid suspensions homogeneously, the nucleation undercooling corresponds to the temperature at which the solidification begins and the released latent heat of solidification will suddenly raise temperature of suspensions. The temperature initially leading to liquid/solid transformation can be recorded by calorimeter, i.e. the nucleation undercooling [28]. Here, the time-temperature curves were recorded by a Yokogawa LR 4110 temperature recorder. The temperature of the recorder was calibrated by a standard platinum resistance thermometer with an uncertainty of ±0.1 $^0$C [24, 29]. The suspensions were put into a hydrophobic plastic tube of 1 ml together with the probe of the temperature recorder. The hydrophobic plastic tube can inhibit heterogeneous nucleation of ice on its surface.

Figure 3 shows the cooling curve of initial suspensions with the volume fraction of $\phi_0$, under the average cooling rate of Rc=5.29×10$^{-2}$ K/s which is in the same order of magnitude with G×V (=1.16×10$^{-2}$ K/s). With the decrease of ambient temperature, the temperature of initial suspensions decreased until the nucleation occurs. A characteristic temperature of -8.4 $^0$C were measured, which also reflects nucleation undercooling of ice (-8.4 $^0$C), as shown in Fig. 3. Since nucleation is random and has probability in a limited range of temperature, multiple measurements of nucleation undercoolings were applied for the identical system. After eight measurements of nucleation undercooling, the average nucleation undercooling for the initial suspension is -6.9±1.8 $^0$C. Moreover, considering the volume fraction of particles in the concentrated layer ahead of the interface during freezing is much higher than $\phi_0$, we also measured the nucleation undercoolings in a dense suspensions with $\phi$=55% (83wt%). Typically, the maximum particle random packing is usually $\phi$=55% [17]. The average undercooling of nucleation for the dense suspensions is gauged as -6.77±1.4 $^0$C for three measurements. The nucleation undercoolings of both initial diluter (-6.9 $^0$C) and denser (-6.77 $^0$C) suspensions are in the same order of magnitude with the data in Ref. [30] (around -12 $^0$C).

The inserted metallic probe of the recorder may also affect the measured ice nucleation in the colloidal suspension. Through measuring the nucleation undercooling of the supernatant centrifuged from the initial suspensions, the effect of the metallic probe on ice nucleation was assessed. The nucleation undercooling of the supernatant was measured as -9.1 $^0$C averagely after seven measurements. The measured nucleation undercoolings of the supernatant (-9.1 $^0$C) and the suspensions (-6.77 $^0$C or -6.9 $^0$C), indicate that the wetting angle of alumina nano-particles is smaller than that of the metallic probe. Therefore, the nucleation gives priority to the help of alumina nano-particles, when the alumina nano-particles and the metallic probe coexist.

Although both the nucleation undercooling and interfacial undercooling have been separately tested in the previous work [18, 28], the rationality of secondary nucleation is never judged by the connections of the nucleation undercooling and the interfacial undercooling. Therefore, comparisons between the nucleation undercooling and the interfacial undercooling of a new ice lens are shown in Fig.4. The reliability of all experimental data in the present paper has been clarified as mentioned above. Both of the nucleation undercoolings of initial diluter (-6.9 $^0$C) and denser (-6.77 $^0$C) suspensions are two orders of magnitude larger than the interfacial undercooling (-0.01 $^0$C) that a new ice lens initially emerges. It reveals that the interfacial undercooling is much smaller than the nucleation undercooling of generation of a new ice lens. Accordingly, secondary nucleation mechanism for ice banding and ice lens can hardly exist.

Furthermore, the definition of nucleation must be clarified, and it is ambiguous used in the previous work [3, 6]. There are distinct meanings of homogeneous nucleation, heterogeneous nucleation and epitaxial growth. Homogeneous nucleation is induced by thermodynamic fluctuations occurring randomly throughout the liquid, uninfluenced by the presence of any extrinsic surfaces, such as internal interfaces provided by dispersed colloidal particles, or through contact with crucible or mold walls needed to support the liquid. Homogeneous nucleation occurs strictly by thermodynamic fluctuations unaided by other effects [31]. The temperature of homogeneous nucleation of water/ice transformation is about -40 $^0$C. In the freezing colloidal suspensions, discussion of homogeneous nucleation is useless in the presence of many particles. Heterogeneous nucleation is influenced by the presence of extrinsic surface of particles. However, undercooling of heterogeneous nucleation is also much larger than the interfacial undercooling as mentioned above. Therefore, the heterogeneous nucleation can hardly exist, i.e., the ice-filled flaw (secondary nucleation), a key part of the theory [6], can hardly exist ahead of the warmest ice lens.

In Ref.[6], new ice lenses can then nucleate from the shrinkage cracks in regions where sufficient geometrical supercooling exists. However, this case is epitaxial growth not nucleation. The major difference is the crystal orientation. Epitaxial growth does not need nucleation but continuously grows from its existing crystalline structure. Meanwhile, epitaxial growth requires almost no supercooling, rather than sufficient geometrical supercooling.

Finally, the magnitude of "geometrical supercooling" proposed in Ref.[6] is largely controversial, which means it can not satisfy the undercooling of secondary nucleation as mentioned above. The small interfacial undercooling for a new ice banding presented here requires a new forming mechanism of ice banding, and should be an essential indicator of the future proposed model. It seems that the frozon fringe mechanism may be the possible choice to explain the ice banding mechanism. However, there is no experimental evidence and with argument. We will try to reexamine the frozon fringe mechanism in the future work.

**Conclusions**

The present paper quantitatively measured the interfacial undercooling that a new ice lens initially emerges and the nucleation undercoolings of initial diluter and denser suspensions. The reliability of all experimental data in the present paper is clarified. The nucleation undercoolings of both initial diluter (-6.9 $^0$C) and denser (-6.77 $^0$C) suspensions are two orders of magnitude larger than the interfacial undercooling (-0.01 $^0$C) that a new ice lens initially emerges. The interfacial undercooling is far from the nucleation undercooling of a new ice lens. Therefore, the secondary nucleation mechanism for ice banding and ice lens can hardly appear. In the future, a new formation mechanism of ice banding should be proposed to explain how a new ice lens forms.

**Acknowledgements**

This research has been supported by Nature Science Foundation of China (Grant Nos. 51371151 and 51571165), Free Research Fund of State Key Laboratory of Solidification Processing (100-QP-2014), the Fund of State Key Laboratory of Solidification Processing in NWPU (13-BZ-2014) and the Fundamental Research

# References

[1] J.A.W. Elliott, S.S.L. Peppin, Physical Review Letters, 107 (2011) 168301.
[2] A.M. Anderson, M.G. Worster, Langmuir, 28 (2012) 16512-16523.
[3] A.M. Anderson, M. Grae Worster, Journal of Fluid Mechanics, 758 (2014) 786-808.
[4] S.S.L. Peppin, R.W. Style, Vadose Zone Journal, 12 (2013).
[5] R.W. Style, S.S. Peppin, Journal of Fluid Mechanics, 692 (2012) 482-498.
[6] R.W. Style, S.S.L. Peppin, A.C.F. Cocks, J.S. Wettlaufer, Physical Review E, 84 (2011) 041402.
[7] A. Lasalle, C. Guizard, E. Maire, J. Adrien, S. Deville, Acta Materialia, 60 (2012) 4594-4603.
[8] S. Deville, Materials, 3 (2010) 1913.
[9] K. Muldrew, J.P. Acker, J. Elliott, L.E. McGann, Life in the frozen state, (2004) 67-108.
[10] S. Taber, The Journal of Geology, 37 (1929) 428-461.
[11] S. Taber, The Journal of Geology, 38 (1930) 303-317.
[12] K. O'Neill, R.D. Miller, Water Resources Research, 21 (1985) 281-296.
[13] A.C. Fowler, SIAM Journal on Applied Mathematics, 49 (1989) 991-1008.
[14] A.W. Rempel, J. Wettlaufer, M. Worster, Journal of Fluid Mechanics, 498 (2004) 227-244.
[15] K. Watanabe, M. Mizoguchi, Journal of Crystal Growth, 213 (2000) 135-140.
[16] S.L. Sobolev, Physics Letters A, 376 (2012) 3563-3566.
[17] S. Deville, E. Maire, G. Bernard-Granger, A. Lasalle, A. Bogner, C. Gauthier, J. Leloup, C. Guizard, Nature Materials, 8 (2009) 966-972.
[18] J. You, L. Wang, Z. Wang, J. Li, J. Wang, X. Lin, W. Huang, arXiv preprint arXiv:1508.02833, (2015).
[19] D. Turnbull, The Journal of chemical physics, 20 (1952) 411-424.
[20] D. Turnbull, Journal of Applied Physics, 21 (1950) 1022-1028.
[21] E.J. Langham, B.J. Mason, Proceedings of the Royal Society of London A: Mathematical, Physical and Engineering Sciences, 247 (1958) 493-504.
[22] D. Turnbull, J.C. Fisher, The Journal of Chemical Physics, 17 (1949) 71-73.
[23] S.C. Brown, D. Payne, Journal of Soil Science, 41 (1990) 547-561.
[24] J. You, L. Wang, Z. Wang, J. Li, J. Wang, X. Lin, W. Huang, Review of Scientific Instruments, 86 (2015) 084901.
[25] S. Peppin, P. Aussillous, H.E. Huppert, M. WORSTER, Journal of Fluid Mechanics, 570 (2007) 69-77.
[26] R. Gilpin, Water Resources Research, 16 (1980) 918-930.
[27] J.M. Konrad, C. Duquennoi, Water Resources Research, 29 (1993) 3109-3124.
[28] T. Kozlowski, Cold Regions Science and Technology, 59 (2009) 25-33.
[29] L. Wang, X. Lin, M. Wang, W. Huang, Journal of Crystal Growth, 406 (2014) 85-93.
[30] S. Deville, E. Maire, A. Lasalle, A. Bogner, C. Gauthier, J. Leloup, C. Guizard, Journal of the American Ceramic Society, 93 (2010) 2507-2510.
[31] M.E. Glicksman, Principles of solidification: an introduction to modern casting and crystal growth concepts, Springer Science & Business Media, 2010.

**List of figures**

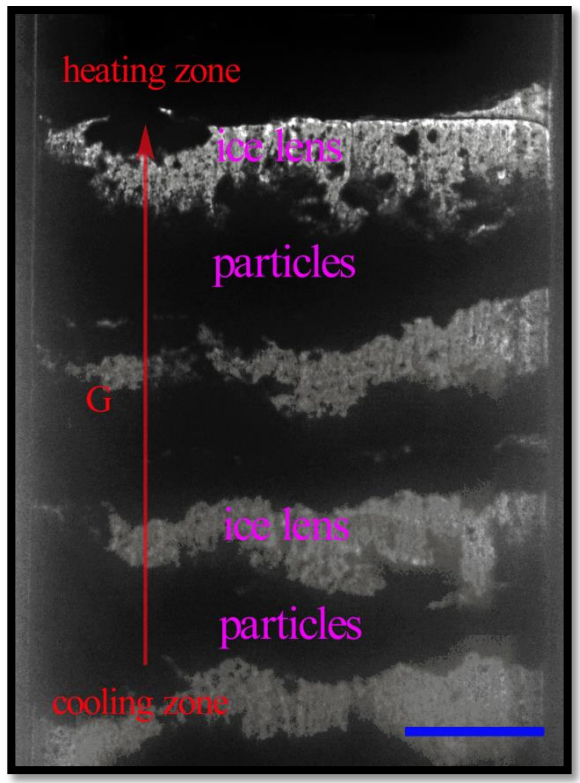

Fig.1 ice banding during freezing alumina suspensions with mean diameter d=50nm, initial volume fraction $\phi_0$=9.74%, temperature gradient G=7.23K/cm and pulling speed V=16μm/s. The scale bar is 200 μm.

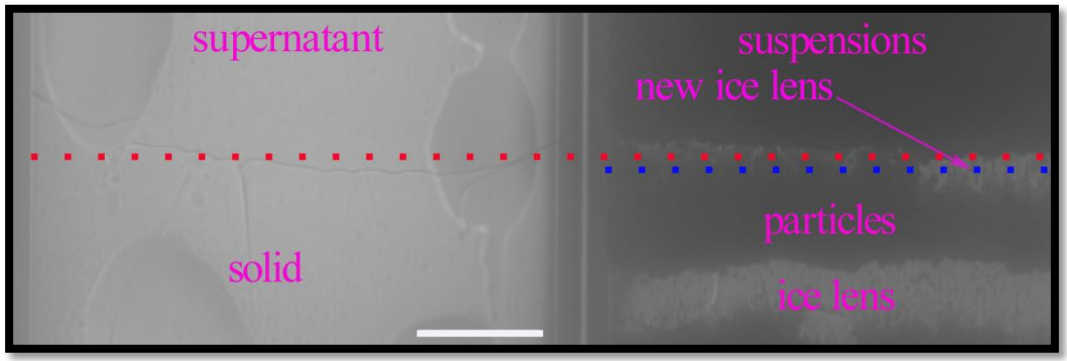

Fig.2 interface positions of supernatant (red dot line) and suspensions (blue dot line). Blue dot line is the position that a new ice lens initially emerges. The scale bar is 200 μm.

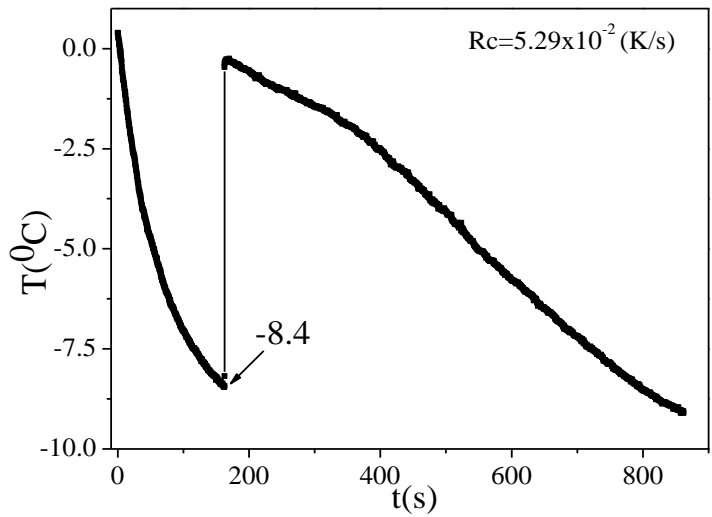

Fig.3 cooling curve of suspensions, under the average cooling rate of Rc=5.29×10$^{-2}$K/s. The saltation of temperature indicates the beginning of solidification.

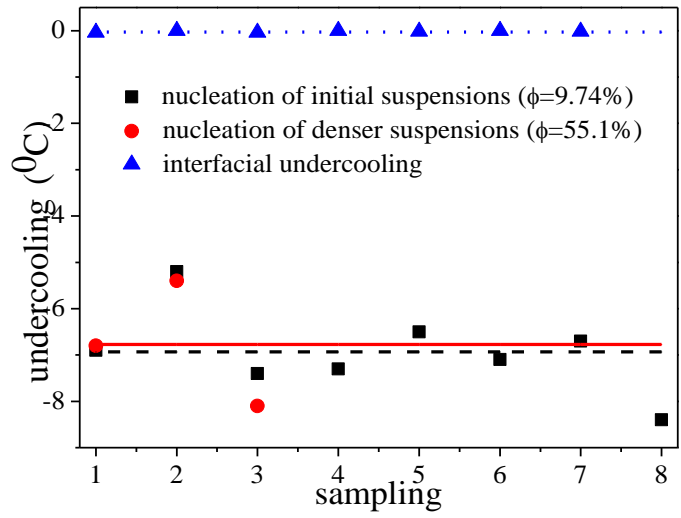

Fig.4 comparison between nucleation undercooling and interfacial undercooling that a new ice lens emerges. The points are experimental data and the lines are average values.